%% file: main.tex
\RequirePackage{etoolbox}
\pdfoutput=1
\documentclass[sigconf,nonacm,bookmarksnumbered,unicode,colorlinks=true,linkcolor=blue,citecolor=blue,urlcolor=blue]{acmart}

\usepackage{amsmath,amsfonts}
\usepackage{algorithmic}
\usepackage{longtable}
\usepackage{tablefootnote}
\usepackage{pifont} 
\usepackage{array} 

\usepackage{subcaption}
\usepackage{geometry}
\usepackage{multicol}
\usepackage{multirow}
\usepackage{listings}
\usepackage{booktabs}
\usepackage{url}
\usepackage{float}
\usepackage{graphicx}
\usepackage{textcomp}
\usepackage{xcolor}
\usepackage{fancyvrb}
\usepackage{tcolorbox}
\usepackage{hyperref}
\usepackage{ulem}
\usepackage{tcolorbox}
\usepackage{array}
\usepackage{longtable}
\usepackage[compatibility=false]{caption}

\definecolor{darkgreen}{rgb}{0.0, 0.5, 0.0}

\lstdefinestyle{customc}{
  language=C,
  frame=single,
  basicstyle=\scriptsize\ttfamily,
  keywordstyle=\color{blue},
  commentstyle=\color{gray},
  numbers=none,
  breaklines=true,
  showstringspaces=false,
  tabsize=2,
}

\newcommand{\greenstars}[1]{\textcolor{darkgreen}{\ding{72}}}
\newcommand{\fullstars}{\greenstars{}\greenstars{}\greenstars{}\greenstars{}} 
\newcommand{\threestars}{\greenstars{}\greenstars{}\greenstars{}} 
\newcommand{\twostars}{\greenstars{}\greenstars{}} 
\newcommand{\onestar}{\greenstars{}} 

\usepackage{color} 
\definecolor{codegreen}{rgb}{0,0.6,0}
\definecolor{codegray}{rgb}{0.5,0.5,0.5}
\definecolor{codepurple}{rgb}{0.58,0,0.82}
\definecolor{backcolour}{rgb}{0.95,0.95,0.92}
\AtBeginDocument{%
  }

\begin{document}

\title{Do Large Language Models Understand Performance Optimization?}

\author{Bowen Cui$^*$}
\email{bcui2@gmu.edu}
\affiliation{%
   \institution{George Mason University}
   \city{Fairfax}
   \state{VA}
   \country{USA}}

\author{Tejas Ramesh$^*$}
\email{tramesh2@gmu.edu}
\affiliation{%
   \institution{George Mason University}
   \city{Fairfax}
   \state{VA}
   \country{USA}}

\author{Oscar Hernandez}
\email{oscar@ornl.gov}
\affiliation{%
   \institution{Oak Ridge National Laboratory}
   \city{Oak Ridge}
   \state{TN}
   \country{USA}}

\author{Keren Zhou}
\email{kzhou6@gmu.edu}
\affiliation{%
   \institution{George Mason University}
   \city{Fairfax}
   \state{VA}
   \country{USA}
}

\thanks{$^*$Equal contribution.}

\begin{abstract}

\input{src/Abstract}

\end{abstract}





\maketitle

\input{src/Introduction}

\input{src/Background}

\input{src/BenchmarkSuite}

\input{src/Evaluation}

\input{src/CaseStudies}


\input{src/Discussion}

\input{src/Conclusions}




\bibliographystyle{plain}
\bibliography{references}

\appendix

\end{document}

%% file: src/Abstract.tex
Large Language Models (LLMs) have emerged as powerful tools for software development tasks such as code completion, translation, and optimization.
However, their ability to generate efficient and correct code, particularly in complex High-Performance Computing (HPC) contexts, has remained underexplored.
To address this gap, this paper presents a comprehensive benchmark suite encompassing multiple critical HPC computational motifs to evaluate the performance of code optimized by state-of-the-art LLMs, including OpenAI o1, Claude-3.5, and Llama-3.2.
In addition to analyzing basic computational kernels, we developed an agent system that integrates LLMs to assess their effectiveness in real HPC applications.
Our evaluation focused on key criteria such as execution time, correctness, and understanding of HPC-specific concepts.
We also compared the results with those achieved using traditional HPC optimization tools.
Based on the findings, we recognized the strengths of LLMs in understanding human instructions and performing automated code transformations.
However, we also identified significant limitations, including their tendency to generate incorrect code and their challenges in comprehending complex control and data flows in sophisticated HPC code.

%% file: src/Introduction.tex
\section{Introduction}

The rise of Large Language Models (LLMs)~\cite{chatgpt-4,claude,gpt-3,Llama3.2} has introduced advanced capabilities to a wide range of applications, such as translation tools~\cite{Zhu2023MultilingualMT}, search engines~\cite{Spatharioti2023ComparingTA}, and recommendation systems~\cite{llm_recommendation}.
Notably, the coding capabilities of LLMs~\cite{Evauating_LLMs_Trained_on_Code,codegeex,codewhisperer} have garnered considerable attention from both academia and industry, particularly in tasks such as code completion~\cite{codellama}, summarization~\cite{Ahmed2022FewshotTL}, and repair~\cite{Xia2022LessTM}.
LLMs are trained on vast datasets~\cite{stack,pile} that cover a diverse range of programming languages and application domains, enabling them to generate code that effectively aligns with the intention conveyed through natural language inputs.
Meanwhile, several studies have proposed benchmarks~\cite{Austin2021ProgramSW,Hendrycks2021MeasuringCC,Evauating_LLMs_Trained_on_Code} for evaluating the capabilities of LLMs in coding.

As LLMs gain widespread popularity, the High-Performance Computing (HPC) community has been assessing and exploring their potential to address various HPC challenges~\cite{HPC,Nichols2024CanLL,Chen2024OMPGPTAG}.
Existing benchmarks predominantly examine whether LLMs can solve problems under human guidance with  example solutions (i.e., multiple shots in-context learning~\cite{Olsson2022IncontextLA}) and tend to evaluate general rather than domain-specific questions.
There is a dearth of studies~\cite{niu2024evaluatingefficiencysourcecode,Qiu2024HowEI} investigating the efficiency (i.e., performance) of LLM-generated code, particularly in the context of complex real-world problems.
Unlike simple benchmarks, HPC code is inherently more complex, requiring not only a deep understanding of hardware, algorithms, and programming languages but also the ability to navigate large code bases and leverage cross-domain knowledge. 
Optimizations such as vectorization, parallelization, loop transformation, and data prefetching demand intricate program analysis and an understanding of multi-level parallelism models to effectively address performance bottlenecks.
Frequently, there exists no straightforward reference solution available for code optimization.

This complexity poses significant challenges for the application of LLMs in HPC tasks. 
Existing models often lack the specialized knowledge and program analysis capabilities necessary to handle these intricacies, making it difficult to achieve the desired level of optimization for HPC applications.
Traditional performance optimization tools~\cite{hpctookit,Shende2006TheTP} have played a pivotal role within the HPC community.
Some of these tools~\cite{MAQAO,KBS-MAQAO,Codee} use static analysis to propose optimizations and occasionally automate code transformations, demonstrating capabilities similar to LLMs.
However, traditional tools face multiple constraints as they rely on heuristic methods, are restricted to limited instances for automatic code transformations, and support only some frontend languages.
The distinctions between traditional optimization tools and LLM inspire our investigation into their impact on HPC benchmarks, with the goal of \textbf{understanding the benefits and limitations of these tools and offering insight into advancing performance optimization by integrating LLMs with traditional performance tools}.




In this research, we curated a benchmark suite comprising 26 representative HPC codes across 11 distinct domains commonly found in HPC problems.
The benchmarks are categorized into three levels based on the complexity and the size of the program.
We evaluated and compared the performance of code optimized by Codee~\cite{Codee} with that optimized by three leading LLMs, OpenAI o1~\cite{GPT-4}, Llama-3.2~\cite{Llama3.2}, and Claude-3.5~\cite{claude}.
We selected Codee as the representative traditional performance tool due to its capability to automate performance analysis and optimization,  provision of in-depth insights into HPC code optimization, and wide adoption~\cite{codee_nersc,richardson2024optimizing} by industry and multiple research institutions.
Using our benchmark suite, we evaluated the capabilities of LLMs and Codee from the following perspectives:
\begin{itemize}
    \item \textbf{Speedups}: Assessing the improvement in execution time achieved through optimization;
    \item \textbf{Scalability (parallelism)}: Measuring the effectiveness of optimizations in scaling across multiple CPU cores and compute nodes;
    \item \textbf{Correctness}: Checking if optimized code outputs the same results as the original code;
    \item \textbf{HPC Commonsense}: Evaluating the models' understanding of domain-specific best practices and principles in HPC;
    \item \textbf{Applicability to Applications}: Investigating the models' ability to handle complex HPC applications
\end{itemize}
In contrast to existing work~\cite{ValeroLara2023ComparingLA, Nichols2024CanLL, Nichols2024PerformanceAlignedLF}, this study is the first to investigate the differences between LLMs and state-of-the-art traditional performance optimization tools.

Moreover, we propose a performance optimization agent~\cite{wang2024survey} for optimizing real HPC applications, integrating insights from both LLMs and traditional performance tools to replicate human optimization of HPC code.
This process typically begins with an existing reference implementation and proceeds with the identification of performance hotspots.
Subsequently, various optimizations are applied to achieve speedup and scalability.
The process is repeated iteratively until the desired speedups are achieved.

The rest of the paper is organized as follows. 
Section~\ref{sec:related work} presents related topics.
Section~\ref{sec:benchmark suite} describes our benchmark suite in detail.
Section~\ref{sec:evaluation} introduces the evaluation workflow and discusses the evaluation results.
Section~\ref{sec:case studies} describes our prototype agent environment and results from applying it to four HPC applications.
We summarize the key findings and rank capabilities of LLMs and Codee in Section~\ref{sec:discussion}.
Section~\ref{sec:conclusions} concludes and presents future directions.



%% file: src/Background.tex
\section{Related Work}
\label{sec:related work}
This section reviews recent studies in related fields to provide background knowledge and highlight the significance of our research.

\paragraph{LLMs for Code}
LLMs can generate code based on human prompts and code snippets. 
Typically, code-specific LLMs are trained on general, large-scale datasets~\cite{jiang2024surveylargelanguagemodels} and further fine-tuned using smaller, code-related datasets. 
Those datasets typically consist of pairs of prompts that represent the purposes of the LLMs, example inputs, and outputs in diff-based formats.
Reinforcement Learning with Human Feedback (RLHF)~\cite{ouyang2022traininglanguagemodelsfollow} enhances LLMs' capability to produce syntactically and functionally correct code, with models such as CodeRL~\cite{NEURIPS2022_8636419d} and PPOCoder~\cite{shojaee2023executionbasedcodegenerationusing} demonstrating significant advancements. 
Furthermore, prompt engineering techniques, including Chain-of-Thought (CoT)~\cite{wei2023chainofthoughtpromptingelicitsreasoning} and Self-Debugging~\cite{chen2023teachinglargelanguagemodels}, facilitate iterative refinements for in-depth code analysis.
Retrieval-augmented generation (RAG) addresses the challenges of repository-level code generation by incorporating cross-file context~\cite{zhang2023repocoderrepositorylevelcodecompletion}. 
Our study evaluates three state-of-the-art LLMs that achieve exceptional results in existing code-related tasks~\cite{lmarena}.

\paragraph{LLM Benchmarks}
A range of benchmarks~\cite{chen2021evaluating,jiang2024surveylargelanguagemodels} has been developed to evaluate the code generation capabilities of LLMs.
Most studies primarily focus on assessing the correctness of the generated code.
Similarity-based metrics such as BLEU~\cite{Papineni2002BleuAM}, ROUGE~\cite{lin-2004-rouge}, and METEOR~\cite{banerjee-lavie-2005-meteor} often fall short in capturing the syntactic and functional correctness of code.
Consequently, execution-based metrics, such as pass@k~\cite{Evauating_LLMs_Trained_on_Code}, have gained traction for their ability to directly evaluate functional validity by running unit tests on generated code.
In addition, there are also studies that capture qualitative aspects like code style and maintainability~\cite{jiang2024surveylargelanguagemodels}.
Existing benchmarks~\cite{nichols2024can,nichols2024performance,niu2024evaluatingefficiencysourcecode,fang2024towards} for evaluating the efficiency of LLMs often focus on narrow aspects, such as their ability to parallelize code, without thoroughly examining their understanding of broader efficiency-related challenges.
Furthermore, prior studies mostly analyze small computational kernels, neglecting larger applications.
In contrast, our study provides a more comprehensive evaluation by assessing the efficiency, correctness, and HPC-specific expertise of LLMs for both small kernels and large applications.

\paragraph{LLM Agents}
An LLM agent~\cite{wang2024survey} is a system designed to utilize LLMs for solving problems by generating a \textit{plan}, executing the plan through external \textit{tools}, gathering feedback from the \textit{environment}, and storing information in \textit{memory} to refine subsequent steps.
It extends the core functionality of LLMs by enabling dynamic interaction with external environments.
Several frameworks~\cite{significantgravitas2023autogpt,wang2024openhands,langchain} have been developed to facilitate the creation of such agents~\cite{cognition2024devin,cursor}.
To enhance solution accuracy, these agents~\cite{zhang2024codeagent} often employ advanced reasoning strategies, such as Chain-of-Thought~\cite{wei2023chainofthoughtpromptingelicitsreasoning}, Tree-of-Thought~\cite{yao2024tree}, and ReAct~\cite{yao2022react}, to determine the most effective invocation strategies.
Additionally, recent research~\cite{kim2023llm,shinn2024reflexion} has explored optimizing the efficiency and accuracy of the agent workflow.
Unlike previous code agents focused on software development, we developed a prototype agent system aimed at evaluating the effectiveness of using LLMs for performance optimization of large applications.

\paragraph{Traditional Performance Analysis Tools}
In the field of HPC, performance analysis tools play a critical role in identifying and resolving bottlenecks to optimize application efficiency across single and multiple compute nodes.
These tools are generally classified into two main categories: dynamic and static tools.
\textit{Dynamic tools}~\cite{hpctookit,shende2006tau} operate at runtime, collecting performance metrics such as execution time, memory usage, and processor utilization and attributing metrics to source files to identify hotspots.
\textit{Static tools}~\cite{djoudi2005maqao,Codee} analyze source code and compilation databases to identify inefficiencies without executing the program.
Recent advancements~\cite{zhou2021automated,zhou2021gpa} have introduced hybrid tools that combine static and dynamic analysis techniques to offer deeper insights.
Traditional performance analysis tools typically rely on heuristics and predefined algorithms to detect performance bottlenecks and inefficiencies.
While this approach ensures precision—static tools apply rigorous algorithms, and dynamic tools capture empirical data from real execution—such methods may lack the flexibility needed to address the complex and evolving challenges in modern HPC environments.
In contrast, LLMs offer enhanced adaptability by modifying code and incorporating additional information through retrieval-augmented generation (RAG).
However, it is crucial to carefully validate the outputs of LLMs to ensure correctness and effectiveness in HPC workflows.

\input{table/table2_cate_datasets}

%% file: table/table2_cate_datasets.tex

\begin{table*}[tp]
\centering
\scriptsize
\caption{Classification of Benchmarks by Computational Motifs}
\label{tab:dataset}
\begin{tabular}{|c|c|c|c|c|p{7cm}|}
\hline
\textbf{Computational Motif} & \textbf{Benchmark} & \textbf{Language} & \textbf{Level} & \textbf{Tokens} &\textbf{Description} \\
\hline
\multirow{5}{*}{Dense Linear Algebra} 
& Durbin~\cite{PolyBenchC-4.2.1} & C & 1 & 538 & Toeplitz system solver, used in signal processing and numerical analysis. \\
& Doitgen~\cite{PolyBenchC-4.2.1} & C & 1 & 667 & Multi-resolution analysis kernel for wavelet transforms. \\
& Cholesky~\cite{PolyBenchC-4.2.1} & C & 1 & 648 & Cholesky decomposition, widely used in dense matrix factorizations. \\
& 2mm~\cite{PolyBenchC-4.2.1} & C & 1 & 795 & Double matrix-matrix multiplications. \\
& Correlation~\cite{PolyBenchC-4.2.1} & C & 1 & 729 & Mean and standard deviation computation for matrix columns. \\
& MATMUL~\cite{codee_performance_demos} & C & 1 & 764 & Basic matrix multiplication. \\
\hline
\multirow{4}{*}{Sparse Linear Algebra} 
& Trisolv~\cite{PolyBenchC-4.2.1} & C & 1 & 623 & Triangular solver, solving sparse triangular systems. \\
& Bicg~\cite{PolyBenchC-4.2.1} & C & 1 & 649 & BiCGStab linear solver, an iterative method for sparse systems. \\
& ATMUX~\cite{codee_performance_demos} & C & 2 & 2185 & Sparse matrix-vector multiplication, often used in physics simulations. \\
& NPB\_CG~\cite{codee_performance_demos} & C & 3 & 20060 & Conjugate gradient solver, an iterative sparse linear algebra algorithm from NAS benchmarks. \\
\hline
Spectral Methods & Deriche~\cite{PolyBenchC-4.2.1} & C & 1 & 864 & Edge detection commonly used in image processing. \\
\hline
\multirow{2}{*}{Monte Carlo} 
& PI~\cite{codee_performance_demos} & C & 1 & 262 & Monte Carlo integration, involves random sampling for numerical simulations. \\
& XSBench~\cite{hpc2021} & C & 3 & 9094 & Monte Carlo Macroscopic Cross Section Lookup Benchmark. \\
\hline
Dynamic Programming & Nussinov~\cite{PolyBenchC-4.2.1} & C & 1 & 2096 & Dynamic programming for RNA secondary structure prediction. \\
\hline
\multirow{4}{*}{Structured Grids} 
& Adi~\cite{PolyBenchC-4.2.1} & C & 1 & 760 & Alternating Direction Implicit method, common in structured grid-based PDEs. \\
& Srad~\cite{srad} & C & 1 & 669 & Speckle reducing anisotropic diffusion for image processing, utilizing structured grids. \\
& Hotspot~\cite{che2009rodinia} & C & 2 & 1384 & 2D structured grid for simulating heat distribution on a chip. \\
& Hotspot3D~\cite{che2009rodinia} & C & 2 & 1117 & 3D heat distribution simulation over a structured grid. \\
\hline
\multirow{3}{*}{N-body Methods} 
& COULOMB~\cite{codee_performance_demos} & C & 2 & 1709 & Coulomb force simulation, typical of N-body problems in molecular dynamics. \\
& Particlefilter~\cite{che2009rodinia} & C & 2 & 2984 & Particle filter, commonly used in N-body filtering methods in robotics and control. \\
& HACCmk~\cite{codee_performance_demos} & C++ & 2 & 3266 & Cosmological simulation using particle-mesh methods for N-body computations. \\
\hline
& Jacobi-1d~\cite{PolyBenchC-4.2.1} & C & 1 & 564 & 1-D Jacobi stencil computation.
\\
Stencils & LBM D2Q37~\cite{hpc2021} & C & 3 & 27364 & Lattice Boltzmann Method (LBM) for fluid dynamics, uses stencil operations for grid updates. \\
\hline
Radiation Transport & Minisweep~\cite{hpc2021} & C & 3 & 10478 & Solves transport equations for radiation, typically used in nuclear engineering. \\
\hline
\end{tabular}
\end{table*}

%% file: src/BenchmarkSuite.tex
\section{Benchmark Suite}
\label{sec:benchmark suite}

In this section, we outline the design rationale behind our benchmark suite, summarized in Table \ref{tab:dataset}.
The suite encompasses a diverse set of  motifs, including Dense and Sparse Linear Algebra, Spectral Methods, Monte Carlo, Dynamic Programming, Structured Grids, N-body Methods, Stencils, and Radiation Transport, which represent the core computational patterns frequently encountered in HPC applications.

Our benchmarks range in scale from small, representative compute kernels, such as basic matrix multiplication (\textit{MATMUL}) and chain matrix multiplication (\textit{2mm}), to large-scale applications like \textit{NPB\_CG} and \textit{LBM D2Q37}.
These benchmarks are categorized into three levels of increasing code size and complexity: level 1 represents the simplest benchmarks, while level 3 encompasses the most complex.
This hierarchical structure enables a comprehensive evaluation of how effectively LLMs and traditional performance tools can tackle foundational computational tasks as well as domain-specific challenges requiring advanced algorithmic and hardware knowledge.
For smaller benchmarks, both LLMs and traditional code analysis tools are capable of analyzing complete code files.
However, for larger applications, LLMs may face output token length limitations, and traditional tools may generate many false-positive suggestions.
To address these challenges, we employ profilers, such as HPCToolkit~\cite{hpctookit}, to identify performance-critical hotspots, focusing optimization efforts on the most computationally intensive code sections.
The benchmarks span multiple programming languages, primarily C and C++, emphasizing the most common coding environments in the HPC community.
To prepare the benchmarks for evaluation, we eliminated any existing OpenMP parallel pragmas.
Additionally, we fully expanded C/C++ macros to avoid potential misinterpretations by LLMs and limitations of traditional tools.

In summary, this multi-level, multi-domain design enables systematic evaluation of the capabilities of LLMs and traditional tools in addressing detailed optimization challenges in small benchmarks and usability issues in large-scale applications.


%% file: src/Evaluation.tex
\section{Evaluation} 
\label{sec:evaluation}

\begin{figure*}[ht] 
  \centering
  \includegraphics[width=0.75\textwidth]{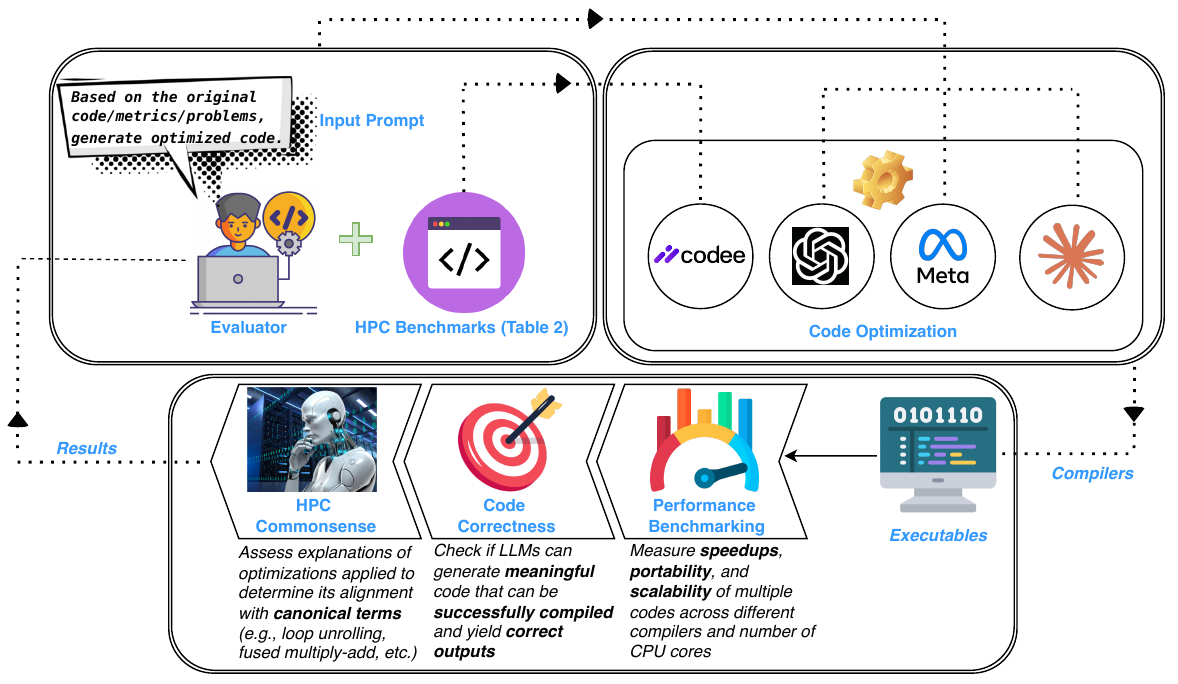}
  \caption{An Overview of our evaluation framework.}
  \label{fig:overview}
\end{figure*}

We describe our evaluation framework as well as a series of experiments conducted using the framework for level 1 and level 2 benchmarks in Figure~\ref{fig:overview}.
It begins by instructing four different tools---Codee, OpenAI o1, Llama-3.2, and Claude-3.5---to optimize existing code written in C/C++.
Note that all these tools accept static code without executing it before providing performance optimization suggestions.

\input{figure/Codee_commands}

Codee incorporates LLVM~\cite{lattner2004llvm} to apply compiler analysis and identify potential optimizations by correlating analysis results with its performance problem catalog~\cite{codee_open_catalog}. 
For each problem, Codee explains why these optimizations are necessary and suggests how they might be implemented.
Some optimizations may be automatically applied by Codee, while others may require manual adjustments.
For example, in Figure~\ref{fig:codee_commands}, we show a sample of how Codee suggests an auto-fix enabled loop interchange optimization for \textit{MATMUL}.

For LLMs, we concatenate initial prompts as shown in Box~\ref{box:prompt_template} with the related code to guide the optimization process.
Note that we do not provide LLMs with any existing optimization examples for a fair comparison with Codee.

The optimized code generated by each tool is compiled into executable binaries, which are then tested with representative inputs to assess \textit{Code Correctness} and \textit{Performance Speedup}.
Moreover, we request each tool to explain the specific optimizations applied, and verify whether the optimization terminology is sensible and consistent with the modifications (i.e., \textit{HPC Commonsense}).

All experiments were performed on a machine equipped with Rocky Linux 8.5 (Green Obsidian) and a single AMD EPYC 7543 32-Core CPU.
The software we used includes GCC/G++ v14.2.0, CLANG/CLANG++ v19.1.5, and Codee v2024.3.5
Table~\ref{tab:models} lists the details of the LLMs we used in the experiments.

\input{table/prompt}

\input{table/table1_models}




\subsection{Experiments Design}

We conducted six experiments to analyze the performance optimization effects of Codee and LLMs.
The details of each experiment are described as follows:

\begin{itemize}
    \item \textbf{Experiment 1 (EX1): Single Serial Optimization.} We apply a single serial optimization that does not include multi-threaded parallelization in this experiment.
    Using Codee, we prefer optimizations that can be auto-fixed or require a simple code change (e.g., modifying a single line or adding a few simple annotations).
    For LLMs, we always instruct them to automatically apply a single optimization.
    The goal is to assess the speedup each tool can achieve through simple optimizations using different compilers.
    \item \textbf{Experiment 2 (EX2): Multiple Serial Optimizations.}
    In this experiment, we select and apply four additional serial optimizations recommended by Codee based on their corresponding importance~\cite{codee_open_catalog}.
    For LLMs, we request them to implement and describe four additional optimizations incrementally.
    Four versions of the optimized code are yielded, and we choose the version that delivers the highest speedup and correct results to indicate the speedup with different compilers.
    The goal is to measure the maximum possible speedups each tool can achieve in the single thread setting with extensive updates.
    \item \textbf{Experiment 3 (EX3): Parallel Optimizations.}
    We instruct Codee and LLMs to consider only parallel optimizations (typically OpenMP multi-threaded parallelism).
    The goal is to check each tool's capability of writing parallel code.
    Note that we separate parallel and serial optimizations to develop a more thorough understanding of the effects of individual optimizations.
    Parallel optimizations can overshadow or interact with serial optimizations, making it difficult to isolate and evaluate the effects of serial optimizations independently.
    \item \textbf{Experiment 4 (EX4): Time Spent on Applying Optimization.}
    We summarize and compare the human time required to invoke tools, prepare prompts, compile and run code, and validate correctness in EX1, EX2, and EX3 for each tool.
    \item \textbf{Experiment 5 (EX5): Correctness.}
    We summarize the correctness of code generated by different tools in EX1, EX2, and EX3. Correctness encompasses the following: checking whether the code is complete, the code compiles without any errors, and the outputs match that of the original code.
    Note that we adopt the \textit{pass@1} metric~\cite{Evauating_LLMs_Trained_on_Code} to measure the correctness of the LLMs, only considering the the top-1 result without prompting the LLMs to produce additional answers.
    \item \textbf{Experiment 6 (EX6): HPC Commonsense.}
    We define a new term-\textit{HPC Commonsense} to understand how the suggested optimizations are relevant to each benchmark and the domain they belong to.
    Codee provides reasons for optimizations during code inspection.
    If an LLM does not provide explanations directly, we prompt it with ``Give me explanations for the optimizations you made.''
\end{itemize}

All experiments were conducted by two graduate students with a limited background in HPC code optimization.
The optimized code and benchmark results have been verified by several experienced performance engineers. 
We compare the execution time for each optimized code generated in EX1, EX2, and EX3 against the original code, with the time averaged over 10 iterations.
For each benchmark, we utilized the built-in input by default and updated certain inputs to ensure that the execution time was sufficiently long.
Note that \textbf{NA} in our figures indicates that the generated code yields incorrect results, fails to compile, or encounters any runtime errors.
For every \textbf{NA} instance, we revert to the original code for measurement of runtime.
We use the following abbreviation for LLMs: O1 (OpenAI o1), L3.2 (Llama-3.2), and C3.5 (Claude-3.5)

\subsection{Experiment 1: Single Serial Optimization}
\label{sub:EX1}

\input{figure/fig2_EX1_EX2}

Figure \ref{fig:EX1_EX2} (a) and (b) present the speedups for each benchmark achieved by optimizing the original code using three different LLMs and Codee, compiled with GCC/G++ and CLANG/CLANG++, respectively. 
The mean speedups achieved by Codee are 1.48$\times$ (GCC/G++) and 1.48$\times$ (CLANG/CLANG++).
In comparison, OpenAI o1 achieves mean speedups of 1.50$\times$ and 1.44$\times$, Llama-3.2 achieves mean speedups of 1.24$\times$ and 1.19$\times$, and Claude-3.5 achieves mean speedups of 1.02$\times$ and 1.04$\times$, using GCC/G++ and CLANG/CLANG++, respectively.
OpenAI o1 typically outperforms Llama-3.2 for speedups and the quantity of correctly produced codes, whereas Claude-3.5 does not reach the same level of efficiency and accuracy.

Due to heuristics and compiler-based code analysis by Codee, we observed no errors in the optimized code.
In contrast, OpenAI o1, Llama-3.2, and Claude-3.5 yield codes that are partially completed or yield incorrect results.
Codee matches or outperforms the LLMs in most of the benchmarks. 
In the case of \textit{MATMUL}, we noticed 2/3 LLMs (OpenAI o1 and Llama-3.2) apply loop interchange as a serial optimization strategy the same as Codee (yielding \textasciitilde 4.5$\times$ speedup). 
We believe the familiarity of LLMs with a widely used algorithm like \textit{MATMUL} during training results in such a speedup.
In another instance, for a more HPC specific algorithm in the \textit{HACCmk} example, out of the LLMs only OpenAI o1 provides significant speedups comparable to Codee (\textasciitilde 5$\times$ speedup), while other LLMs cannot optimize the code well.
Llama-3.2 and Claude-3.5 struggle to achieve significant optimization in level 2 benchmarks, such as \textit{Particlefilter}.
The results with CLANG/CLANG++ are generally consistent with the results obtained with GCC/G++.




\subsection{Experiment 2: Multiple Serial Optimizations}
\label{subsec:EX2}

Figure \ref{fig:EX1_EX2} (c) and (d) present the optimization results for each benchmark after applying multiple serial optimizations.
Codee achieves mean speedups of 1.71$\times$ (GCC/G++) and 1.78$\times$ (CLANG/CLANG++).
OpenAI o1 achieves mean speedups of 2.07$\times$ and 2.13$\times$, Llama-3.2 achieves mean speedups of 1.30$\times$ and 1.32$\times$, and Claude-3.5 achieves mean speedups of 1.31$\times$ and 1.31$\times$.
As EX2 codes are derived from the optimizations implemented in EX1, errors present in EX1 persist and those codes are still identified as \textbf{NA} (e.g., \textit{Srad}).

In most cases, we observed that higher speedups are achieved after applying multiple optimizations.
For example, in \textit{Hotspot3D}, Codee shows a significant improvement in multiple-round optimization compared to single-round optimization, with a speedup of 1.80$\times$ using CLANG/CLANG++.
In an interesting scenario for \textit{Particlefilter}, apart from suggesting regular canonical optimizations, OpenAI o1 goes a step further in suggesting the implementation of binary search as opposed to the existing linear search approach resulting in a speedup of 9.8$\times$ on both the compilers.
This finding highlights LLMs' capability to offer adaptive optimizations from the perspective of algorithmic time complexity.




\subsection{Experiment 3: Parallel Optimization} 
\label{subsec:EX3}
\input{figure/fig3_EX3}

Figure \ref{fig:EX3_GCC} and Figure \ref{fig:EX3_CLANG} demonstrate the speedups achieved by parallel optimizations using each tool with 4 to 32 threads, compiled with GCC/G++ and CLANG/CLANG++ respectively.
For GCC/G++, the mean speedups of Codee, OpenAI o1, Llama-3.2, and Claude-3.5 across 4 to 32 threads are 4.75$\times$, 5.01$\times$, 2.89$\times$, and 3.38$\times$, respectively.
For CLANG/CLANG++, the mean speedups of Codee, OpenAI o1, Llama-3.2, and Claude-3.5 across 4 to 32 threads are 4.12$\times$, 5.04$\times$, 2.46$\times$, and 3.64$\times$, respectively.
OpenAI o1 demonstrates the best performance in recommending strategies for parallel optimization across different compilers.
Codee does suggest good parallel optimizations as well, offering auto-fixes for most of the commonly used optimization strategies.
Notably, in \textit{HACCmk} and \textit{Hotspot3D}, Codee's performance is particularly impressive, achieving 40.00$\times$ and 21.20$\times$ at 32 threads respectively.
As for the LLMs, it should be noted that Claude-3.5 fails to optimize 8 out of 20 benchmarks.

While the LLMs are good at yielding efficient optimizations for \textit{MATMUL} as discussed in previous sections, in the case of parallelization, all of them fail to suggest meaningful optimizations that yield speedups.
Interestingly, while the OpenAI o1 yields code with remarkable speedups for \textit{MATMUL}, it applies serial optimizations instead of using OpenMP pragmas, suggesting that it does not follow users' instructions and breaks the strictly parallel optimization requirement of our experiment.
Llama-3.2 and Claude-3.5 end up not only parallelizing the matrix multiplication function, but also parallelizing the outer loop that benchmarks the MATMUL function multiple times, resulting in complex nested parallelism and excessive memory contention.
In the case of \textit{Hotspot3D}, Codee recommends using \texttt{\#pragma omp for schedule(auto)} to achieve a higher speedup than \texttt{\#pragma omp parallel for} suggested by the LLMs.
A similar scenario also occurs in \textit{HACCmk}.

Interestingly with \textit{Durbin} and \textit{Jacobi-1d}, the performance declines with an increase in thread counts.
This occurs as a result of applying OpenMP pragmas to the inner loop instead of the outer loop of the nested loops present in the program.
When parallelizing the inner loop, the workload per thread might be too small relative to the overhead of creating and synchronizing threads.
Another interesting scenario occurs with the case of \textit{Srad} where Llama-3.2 is the only model that failed to scale using the CLANG/CLANG++ compiler, as shown in Figure~\ref{fig:EX3_CLANG}.
However, its GCC/G++ counterpart does achieve non-trivial speedups.
Upon investigation, we found that  GCC/G++ automatically fused nested loops for better scheduling. 
While CLANG/CLANG++, on the other hand, only parallelizes the outmost loop with \texttt{\#pragma omp parallel for}.

\subsection{Experiment 4: Time Spent on Applying Optimizations}
\label{subsec:EX4}

\input{table/table5_EX4}

Table \ref{tab:Opt-time} summarizes the time required to apply optimizations using each tool, including activities such as prompting the model, waiting for responses, implementing necessary code changes, and validating the results.
Codee includes built-in auto-fixes for specific optimizations, allowing it to automatically transform code regardless of its size.
Although many optimizations still require manual adjustments to the code, Codee provides the file, function, and line number where the optimizations should be applied.
In comparison, LLMs can often automate code transformations.
Nonetheless, Codee is generally faster than LLMs across experiments for the following reasons in EX2 and EX3.
First, since LLMs are invoked via remote servers and the models used are large, the process of generating responses can be slow, especially when multiple sequential optimizations are applied, resulting in lengthy outputs.
Notably, querying the Llama-3.2 API via Together AI~\cite{together_ai} introduces a significant delay compared to other language models.
Furthermore, LLMs are typically limited by output token constraints (e.g., 4k tokens for OpenAI o1), which can prevent them from generating complete code outputs, especially in EX2 with multiple serial optimizations.
Moreover, some LLM outputs contain extra text or incomplete code, necessitating the manual extraction of code from LLM outputs.
It is interesting to observe that OpenAI o1 often produces clean code despite its longest inference time, thereby decreasing the time required for subsequent processing.

%

\subsection{Experiment 5: Correctness}
\label{subsec:EX5}

\input{table/table3_correctness}

Table~\ref{tab:correctness} shows that Codee consistently demonstrates a perfect correctness rate across all benchmarks, outperforming the LLMs, which do not tend to always provide the desired results.
Codee can accurately optimize code across various scenarios using compiler-based analysis.
Despite their broader applicability, LLMs often struggle with analyzing data flow, cross-iteration dependencies, and function invocations, particularly in complex benchmarks.

In EX1, all LLMs fail to yield codes that match expected outputs/results for the \textit{Srad} benchmark.
As shown in Figure~\ref{fig:original-code}, the correct logic is to first compute all derivatives from the old \texttt{J}, then apply them in a separate loop.
OpenAI o1 fails because it incorrectly calculates the directional derivatives---\texttt{dN}, \texttt{dS}, \texttt{dW}, and \texttt{dE} using the image \texttt{J} in the same loop that updates \texttt{J}.
While traversing \texttt{J} in the row-major order, some positions (those with smaller indices) are already updated while others are not, leading to old and new values being mixed, and this breaks the fundamental requirement of the \textit{Srad} algorithm that the directional derivatives for each pixel must be computed from the same iteration.
Llama-3.2 and Claude-3.5 fail due to similar reasons by collapsing two nested loops into one.

\input{figure/Srad_minted}

LLMs also tend to output code that is completely irrelevant for the given context despite giving well-structured prompts as inputs.
For the \textit{COULOMB} example in EX2, Llama-3.2 produces some code that is completely unrelated to the prompt which provided a code snippet that needed to be optimized, which is a classic case of \textit{Hallucination} \cite{xu2024hallucinationinevitableinnatelimitation}.
In the case of \textit{MATMUL} in EX3, when instructed to apply parallel optimizations, OpenAI o1 tends to change the memory access patterns, which isn't really a parallel optimization strategy, an example of LLMs not following instructions to yield codes that aren't meaningful.
LLMs also lack a precise understanding of compiler features. 
In EX3, for \textit{Cholesky}, Claude-3.5 fails to generate code that yields correct results. 





\subsection{Experiment 6: HPC Commonsense}
\label{subsec:EX6}
\input{figure/Domains}

\input{figure/optimization_summary}

We assess HPC commonsense of tools from two perspectives.
First, we assess if the speedup achieved by each tool across all computation motifs in Table~\ref{tab:dataset}.
Second, we analyze the distribution of optimizations applied by each tool, as well as the rationale behind the optimizations.

From Figure~\ref{subfig:ex1_ex2_domain} and Figure~\ref{subfig:ex3_domain}, we observe that different domains demonstrate varying speedups when employing specific tools.
Codee achieves good results in Spectral Methods and Structured Grids using serial optimization strategies, while excelling in Dense Linear Algebra and N-body Methods in a parallel setting.
We believe Codee's familiarity with the nature of computation logic involved in these domains makes it a better code optimizer in such use cases.
For instance, from Figure~\ref{fig:EX6} it is quite evident that a large chunk of the optimizations suggested by Codee revolve around mathematical optimizations including ``Fused multiply-add,'' ``Change of precision'', and ``Mathematical simplification.'' Codee, on the other hand, fails to yield speedups in the Dynamic Programming domain, especially in the parallel setting.
In comparison, OpenAI o1 achieves a speedup of \textasciitilde7$\times$ in the Dynamic Programming domain in EX3.
For Dynamic Programming problems, the original form of the many state transition equations cannot be parallelized, due to interdependencies between successive iterations. 
However, we found that o1 can change the access order of the state transition table to be populated, reducing dependencies so that each column in the table can be updated by individual threads~\cite{venkatachalam2014faster}.
This finding highlights the potential of LLMs in understanding user code to perform adaptive optimizations.


Figure~\ref{fig:EX6} illustrates the distribution of optimizations applied by each tool.
Codee's optimizations are straightforward to summarize, as it categorizes inefficiencies using canonical terminologies such as ``fused multiply-add,'' ``loop fission,'' and ``loop fusion,'' following the structured taxonomy of performance issues outlined in \cite{codee2024opencatalog}.
Among these, ``fused multiply-add'' and ``OpenMP scoping'' are the most frequently applied optimizations.
In contrast, LLMs propose a broader spectrum of optimizations, characterized by more flexible transformations.
These are often described using general terms such as ``pre-computing constants'' and ``reducing function overhead,'' which, while versatile, may lack the specificity required to precisely identify the transformations applied.
A notable trend among LLMs is the frequent emphasis on memory optimizations, which constitute a significant portion of their recommendations, as reflected in the distribution chart.
For scenarios involving nested loop structures, OpenAI o1, like Codee, commonly suggests ``loop reordering'' as an optimization strategy.
However, this opportunity is not identified by other models, such as Llama-3.2 and Claude-3.5.
When analyzing the code provided in the initial prompt, OpenAI o1 distinguishes itself by offering detailed logical reasoning to justify the chosen optimizations.
In contrast, other LLMs tend to lack this level of explanation.
Codee also generates a comprehensive report of the entire codebase, including line numbers from all dependent files, performance problems with the code, and potential optimizations.
However, it is important to note that Codee's suggestions are based on predefined heuristics and cannot customize based on application-specific characteristics.

%% file: figure/Codee_commands.tex
\begin{figure}[tp]
    \centering
    \captionsetup{skip=3pt}
\begin{lstlisting}[
        basicstyle=\ttfamily\tiny, % Adjust font style and size
        label={lst:codee_commands}
        frame=single
    ]
codee checks --verbose  --target-arch cpu -- gcc -c main.c -I include/ -O3
Date: 2025-01-09 Codee version: 2024.4.2 License type: Full
Compiler invocation: gcc -c main.c -I include/ -O3
[1/1] main.c ... Done
CHECKS REPORT
main.c:16:9 [PWR039] (level: L1): Consider loop interchange to improve the locality of reference and enable vectorization
  Loops to interchange:
    16:         for (size_t j = 0; j < n; j++) {
    17:             for (size_t k = 0; k < p; k++) {
  Suggestion: Interchange inner and outer loops in the loop nest to improve performance
  Documentation: https://github.com/codee-com/open-catalog/tree/main/Checks/PWR039
  AutoFix: codee rewrite --memory loop-interchange --in-place main.c:16:9 -- gcc -c main.c -I include/ -O3
\end{lstlisting}
    \caption{An example of Codee suggesting optimizations for \textit{MATMUL}}
    \label{fig:codee_commands}
\end{figure}

%% file: table/prompt.tex
\begin{tcolorbox}[colback=gray!5!white, colframe=black, title={\footnotesize Prompt template for EX1 \& EX2 \& EX3}., label={box:prompt_template}]
\footnotesize
\textbf{\# System Prompt (Common for all experiments and LLMs)} \\
You are a code generation/optimization assistant. Given a prompt your output must only be a compilable source code. The computation environment is a Linux system (Rocky Linux 8.5 Green Obsidian) and a single AMD EPYC 7543 32-Core CPU. The C/C++ language compilers available are: GCC/G++ v14.2.0 and CLANG/CLANG++ v19.1.5\\
\textbf{\# Prompt for EX1} \\
Provide the C/C++ code with a single serial optimization without removing any of the existing functions or header files and without adding any new functions or print statements.\\
\textbf{\# Prompt for EX2} \\
Propose an additional serial optimization that can be applied without removing any of the existing functions or header files and without adding any new functions or print statements.\\
\textbf{\# Prompt for EX3} \\
Based on the original code, provide optimized parallel C/C++ code  without removing any of the existing functions or header files and  without adding any new functions or print statements.

\end{tcolorbox}

%% file: table/table1_models.tex
\begin{table}[ht!]
\scriptsize
\centering
\caption{Comparison of LLMs\tablefootnote{``Context Window'' here refers to the total input + output token limit of the model.}}
\begin{tabular}{|c|c|c|c|c|}
\hline
\textbf{Model} & \textbf{Context Window} & \textbf{Max Output} & \textbf{\#Parameters} & \textbf{Publish Date}\\
\hline
OpenAI o1~\cite{GPT-4} & 128k tokens & 4096 tokens & $\sim$ & Dec 5, 2024\\
\hline
Llama-3.2~\cite{Llama3.2} & 128k tokens & 4096 tokens & 90B & Sep 25, 2024\\
\hline
Claude-3.5~\cite{claude} & 200k tokens & 8192 tokens & $\sim$ & June 20, 2024\\
\hline
\end{tabular}
\label{tab:models}
\end{table}

%% file: figure/fig2_EX1_EX2.tex
\begin{figure*}[ht]
\centering
  \includegraphics[width=1\linewidth]{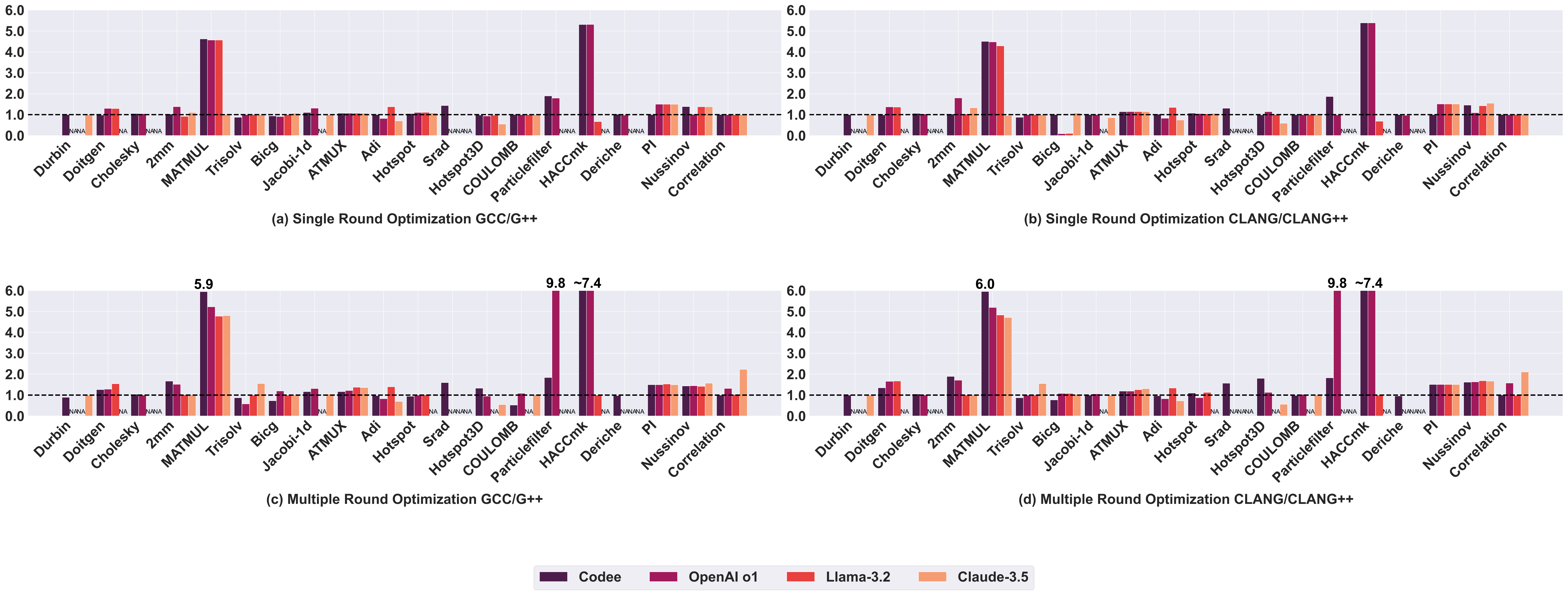}
  \caption{Speedups of single and multiple round serial optimizations. The Y axis indicates the speedup and the horizontal lines denote a speedup of 1.0. Benchmarks that achieved speedup >=5.9 in multiple round optimization across GCC/G++ and CLANG/CLANG++ compilers have been marked above the respective bars.}
  
\label{fig:EX1_EX2}
\vspace{-1.0em}
\end{figure*}

%% file: figure/fig3_EX3.tex
\begin{figure}[tp]
\centering
\includegraphics[width=1\linewidth]{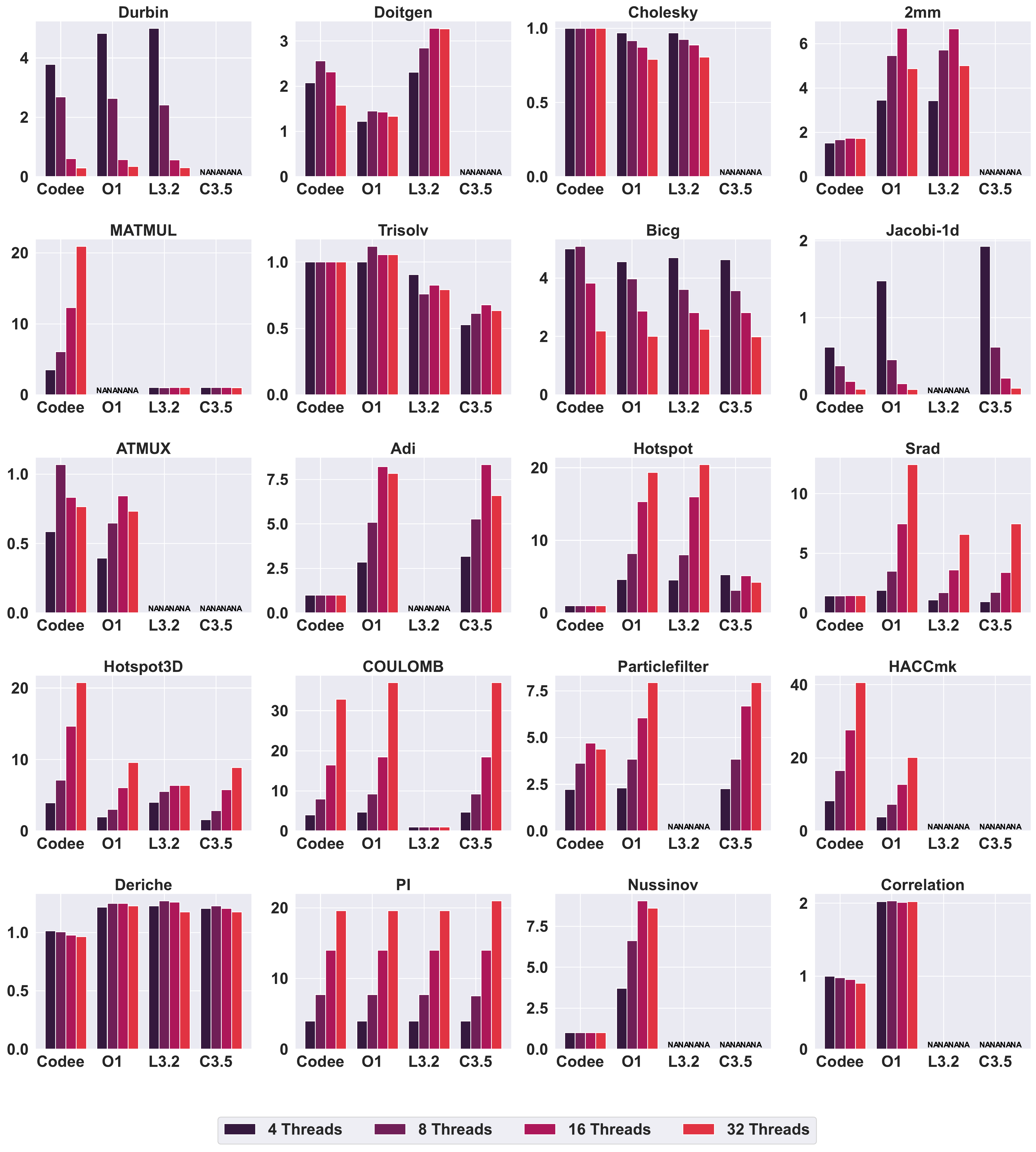}
\caption{Results from parallel optimizations using GCC/G++ compiler. The Y axis indicates the speedup.}
\label{fig:EX3_GCC}
\vspace{-1em}
\end{figure}

\begin{figure}[tp]
\centering
\includegraphics[width=1\linewidth]{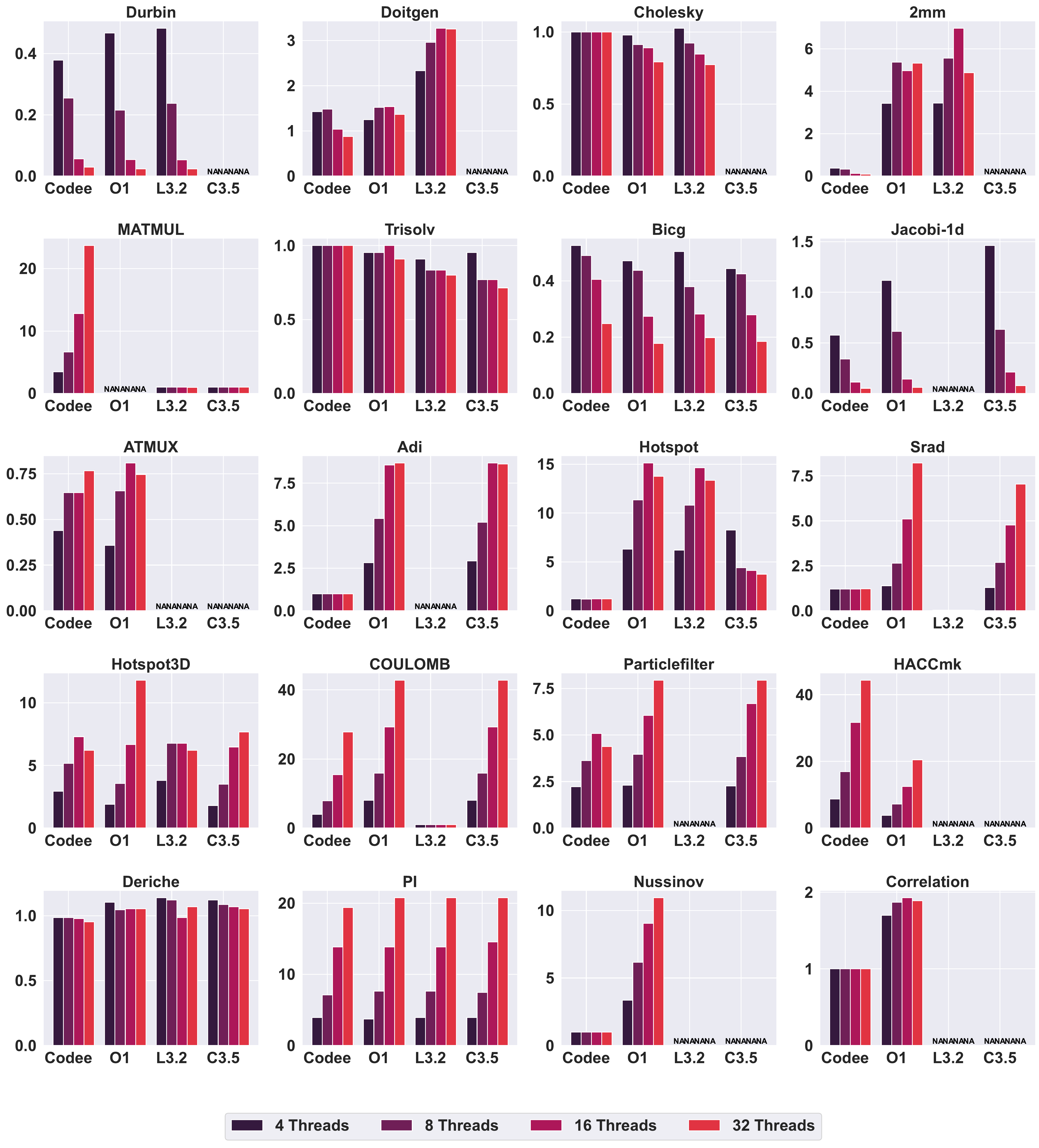}
\caption{Results from parallel optimizations using CLANG/CLANG++ compiler.
The Y axis indicates the speedup.}
\label{fig:EX3_CLANG}
\vspace{-1em}
\end{figure}

%% file: table/table5_EX4.tex

\begin{table}[!t]
\scriptsize
\centering
\caption{Time taken to apply code optimizations}
\label{tab:Opt-time}
\begin{tabular}{|c|ccc|}
\hline
\multirow{2}{*}{Tool} & \multicolumn{3}{c|}{Average Time (Per Benchmark)}                                                                     \\ \cline{2-4} 
                      & \multicolumn{1}{c|}{EX1} & \multicolumn{1}{c|}{EX2}                   & EX3 \\ \hline
Codee                 & \multicolumn{1}{c|}{$\sim$3 mins}  & \multicolumn{1}{l|}{$\sim$4 mins}                      & $\sim$2 mins            \\ \hline
OpenAI o1            & \multicolumn{1}{l|}{$\sim$2 mins}  & \multicolumn{1}{l|}{$\sim$6 mins ($\sim$1.5 mins/trial)} & $\sim$5 mins            \\ \hline
Llama-3.2                 & \multicolumn{1}{l|}{$\sim$2 mins}  & \multicolumn{1}{l|}{$\sim$9 mins ($\sim$2.25 mins/trial)} & $\sim$5 mins            \\ \hline
Claude-3.5              & \multicolumn{1}{l|}{$\sim$2 mins}  & \multicolumn{1}{l|}{$\sim$6 mins ($\sim$1.5 mins/trial)} & $\sim$4 mins            \\ \hline
\end{tabular}
\end{table}

%% file: table/table3_correctness.tex
\begin{table*}[tp]
\scriptsize
\centering
\caption{Comparison of correctness of code optimized by Codee and LLMs in EX1, EX2, and EX3.}
\begin{tabular}{|l|cccc|cccc|cccc|l|}
\hline
\multirow{2}{*}{\textbf{Outcome}} & \multicolumn{4}{c|}{\textbf{EX1}}                                               & \multicolumn{4}{c|}{\textbf{EX2}}                                               & \multicolumn{4}{c|}{\textbf{EX3}}                                               & \multirow{2}{*}{\textbf{Total}} \\ \cline{2-13}
 & \multicolumn{1}{c|}{\textbf{Codee}} 
 & \multicolumn{1}{c|}{\textbf{O1}} 
 & \multicolumn{1}{c|}{\textbf{L3.2}} 
 & \textbf{C3.5} 
 & \multicolumn{1}{c|}{\textbf{Codee}} 
 & \multicolumn{1}{c|}{\textbf{O1}} 
 & \multicolumn{1}{c|}{\textbf{L3.2}} 
 & \textbf{C3.5}
 & \multicolumn{1}{c|}{\textbf{Codee}} 
 & \multicolumn{1}{c|}{\textbf{O1}} 
 & \multicolumn{1}{c|}{\textbf{L3.2}} 
 & \textbf{C3.5} 
 &  \\ \hline
\textbf{Incorrect} & \multicolumn{1}{c|}{0}  & \multicolumn{1}{c|}{2}  & \multicolumn{1}{c|}{6}  &  6 & \multicolumn{1}{c|}{0}  & \multicolumn{1}{c|}{3}  & \multicolumn{1}{c|}{8}  & 7  & \multicolumn{1}{c|}{0}  & \multicolumn{1}{c|}{1}  & \multicolumn{1}{c|}{7}  & 8  & 48 \\ \hline
\ \ \ Compilation errors         & \multicolumn{1}{c|}{0}  & \multicolumn{1}{c|}{0}  & \multicolumn{1}{c|}{0}  &  0 & \multicolumn{1}{c|}{0}  & \multicolumn{1}{c|}{0}  & \multicolumn{1}{c|}{0}  & 0  & \multicolumn{1}{c|}{0}  & \multicolumn{1}{c|}{0}  & \multicolumn{1}{c|}{1}  & 2  & 3 \\ \cline{2-14}
\ \ \ No LLM generated code      & \multicolumn{1}{c|}{0}  & \multicolumn{1}{c|}{1}  & \multicolumn{1}{c|}{3}  &  2 & \multicolumn{1}{c|}{0}  & \multicolumn{1}{c|}{1}  & \multicolumn{1}{c|}{6}  & 5  & \multicolumn{1}{c|}{0}  & \multicolumn{1}{c|}{0}  & \multicolumn{1}{c|}{1}  & 0  & 19 \\ \cline{2-14}
\ \ \ Incorrect results - Output mismatch & \multicolumn{1}{c|}{0}  & \multicolumn{1}{c|}{1}  & \multicolumn{1}{c|}{3}  &  3 & \multicolumn{1}{c|}{0}  & \multicolumn{1}{c|}{2}  & \multicolumn{1}{c|}{2}  & 2  & \multicolumn{1}{c|}{0}  & \multicolumn{1}{c|}{0}  & \multicolumn{1}{c|}{5}  & 6  & 24 \\ \cline{2-14}
\ \ \ LLM - Failed to follow instructions & \multicolumn{1}{c|}{0}  & \multicolumn{1}{c|}{0}  & \multicolumn{1}{c|}{0}  &  1 & \multicolumn{1}{c|}{0}  & \multicolumn{1}{c|}{0}  & \multicolumn{1}{c|}{0}  & 0  & \multicolumn{1}{c|}{0}  & \multicolumn{1}{c|}{1}  & \multicolumn{1}{c|}{0}  & 0  & 2 \\ \hline
\textbf{Correct}                 & \multicolumn{1}{c|}{20} & \multicolumn{1}{c|}{18} & \multicolumn{1}{c|}{14} & 14 & \multicolumn{1}{c|}{20} & \multicolumn{1}{c|}{17} & \multicolumn{1}{c|}{12} & 13 & \multicolumn{1}{c|}{20} & \multicolumn{1}{c|}{19} & \multicolumn{1}{c|}{13} & 12 & 192 \\ \hline
\textbf{Total}                   & \multicolumn{1}{c|}{20} & \multicolumn{1}{c|}{20} & \multicolumn{1}{c|}{20} & 20 & \multicolumn{1}{c|}{20} & \multicolumn{1}{c|}{20} & \multicolumn{1}{c|}{20} & 20 & \multicolumn{1}{c|}{20} & \multicolumn{1}{c|}{20} & \multicolumn{1}{c|}{20} & 20 & 240 \\ \hline
\end{tabular}
\label{tab:correctness}
\end{table*}

%% file: figure/Srad_minted.tex
\begin{figure}[t]
    \centering
    \begin{lstlisting}[style=customc]
for (int i = 0; i < rows; i++)
    for (int j = 0; j < cols; j++) {
        // Calculate directional derivatives
        dN[k] = J[iN[i] * cols + j] - Jc;
        dS[k] = J[iS[i] * cols + j] - Jc;
        dW[k] = J[i * cols + jW[j]] - Jc;
        dE[k] = J[i * cols + jE[j]] - Jc;
    }
for (int i = 0; i < rows; i++)
    for (int j = 0; j < cols; j++) {
        // Load dN, dS, dW, and dE
        D = cN * dN + cS * dS + cW * dW + cE * dE;
        // Update J
        J[k] = J[k] + 0.25 * lambda * D;
    }
    \end{lstlisting}
    \caption{Original code snippet of \textit{Srad}}
    \label{fig:original-code}
\end{figure}

\begin{figure}[t]
    \centering
    \begin{lstlisting}[style=customc]
for (int i = 0; i < rows; i++) {
    for (int j = 0; j < cols; j++) {
        // Calculates directional derivatives dN, dS, dW, dE based on J
        D = cN * dN + cS * dS + cW * dW + cE * dE;
        J[k] = J[k] + 0.25 * lambda * D;
    }
}
    \end{lstlisting}
    \caption{Optimized code of \textit{Srad} - OpenAI o1}
    \label{fig:optimized-code-o1}
\end{figure}

%% file: figure/Domains.tex
\begin{figure}[t]
    \centering
    \begin{minipage}[b]{1.0\linewidth}
        \centering
        \captionsetup{skip=3pt}
        \includegraphics[width=1.0\linewidth]{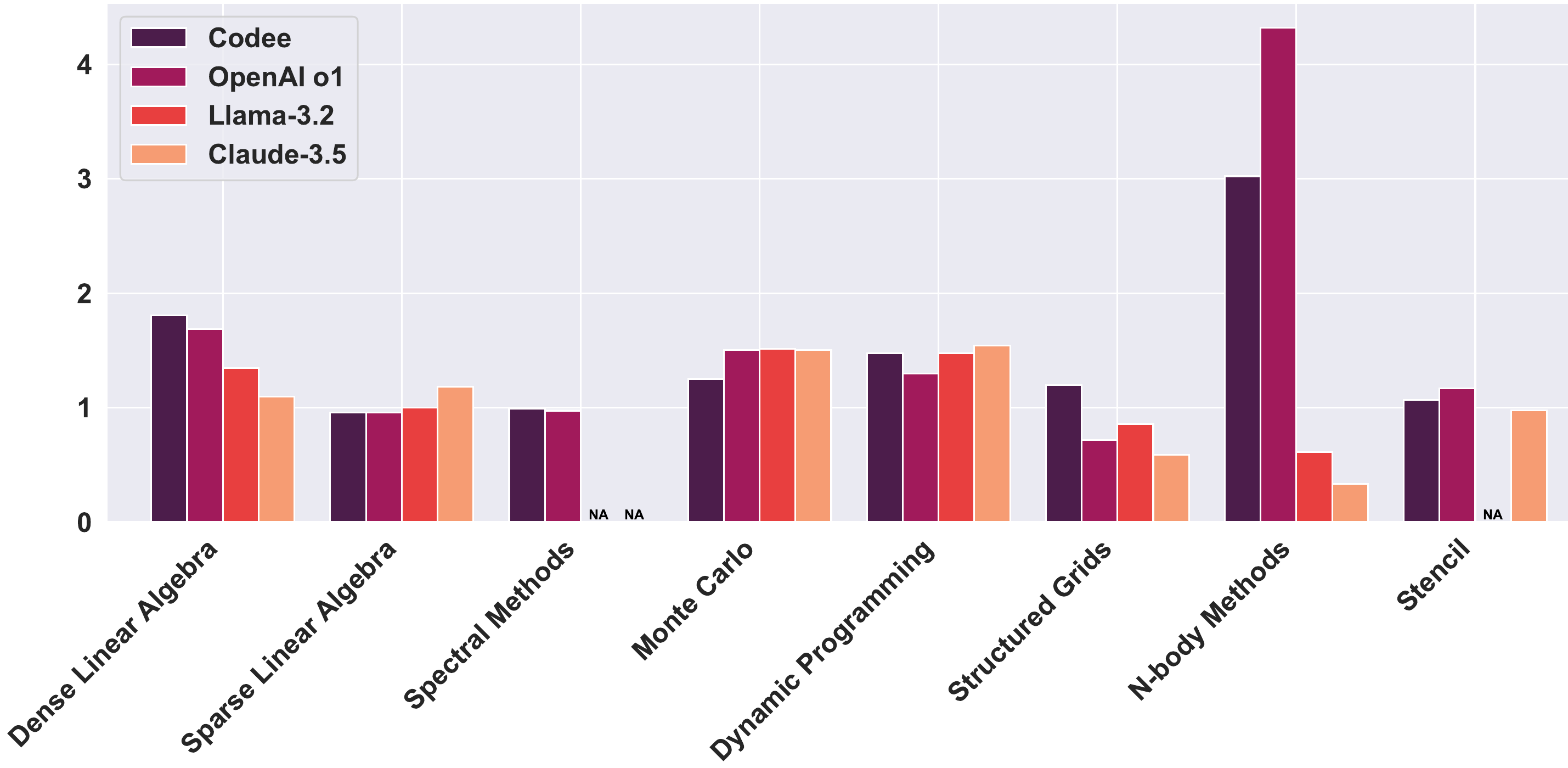} 
        \captionof{figure}{Average speedups (Y-axis) achieved by the tools by domain in EX1 and EX2.}
        \label{subfig:ex1_ex2_domain}
    \end{minipage}
    \hspace{0.02\textwidth} 
    \begin{minipage}[b]{1.0\linewidth}
        \centering
        \captionsetup{skip=3pt}
        \includegraphics[width=1.0\linewidth]{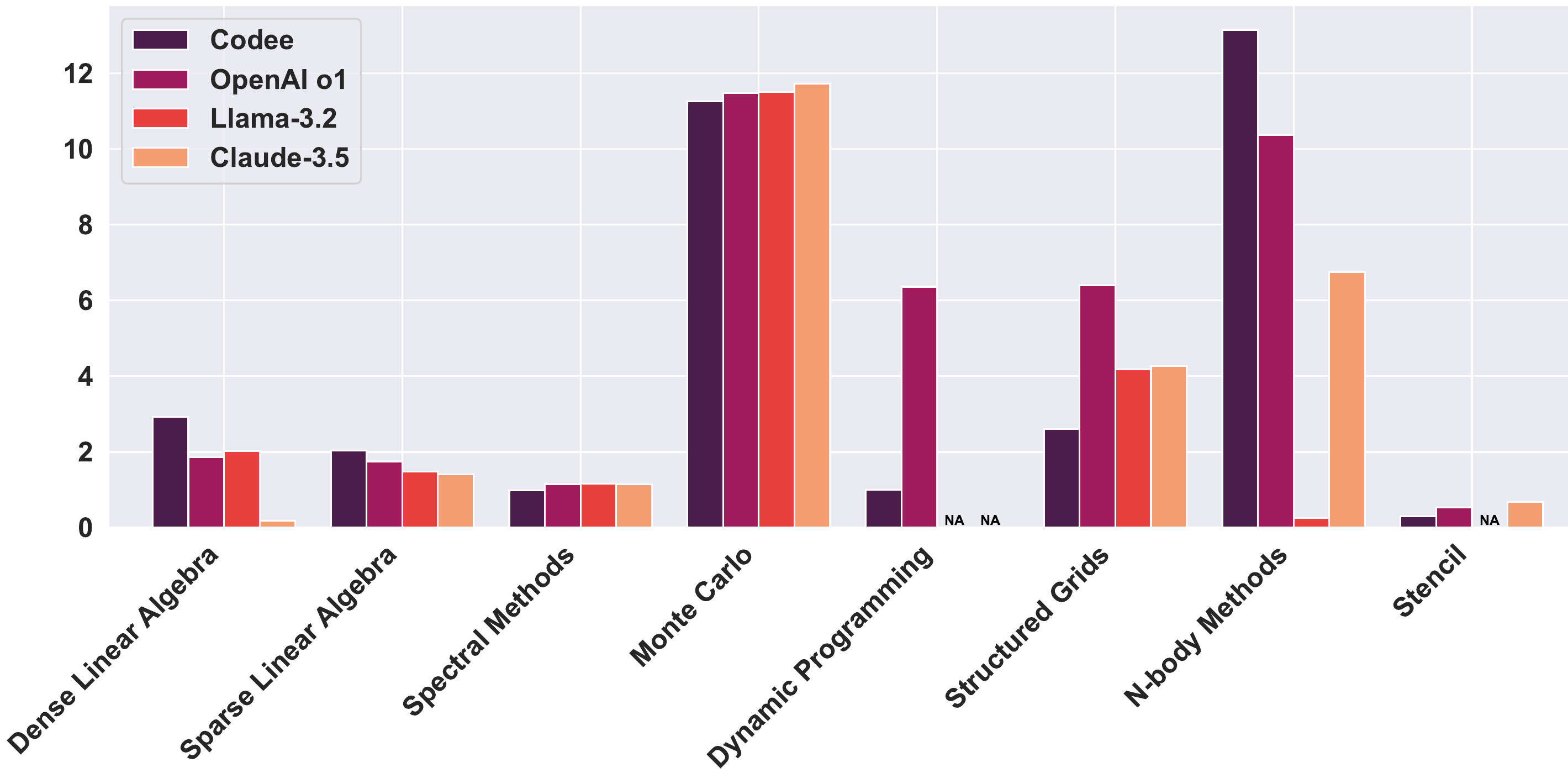} 
        \captionof{figure}{Average speedups (Y-axis) achieved by the tools by domain in EX3.}
        \label{subfig:ex3_domain}
    \end{minipage}
\end{figure}

%% file: figure/optimization_summary.tex
\begin{figure}[ht]
\centering
  \includegraphics[width=0.8\linewidth]{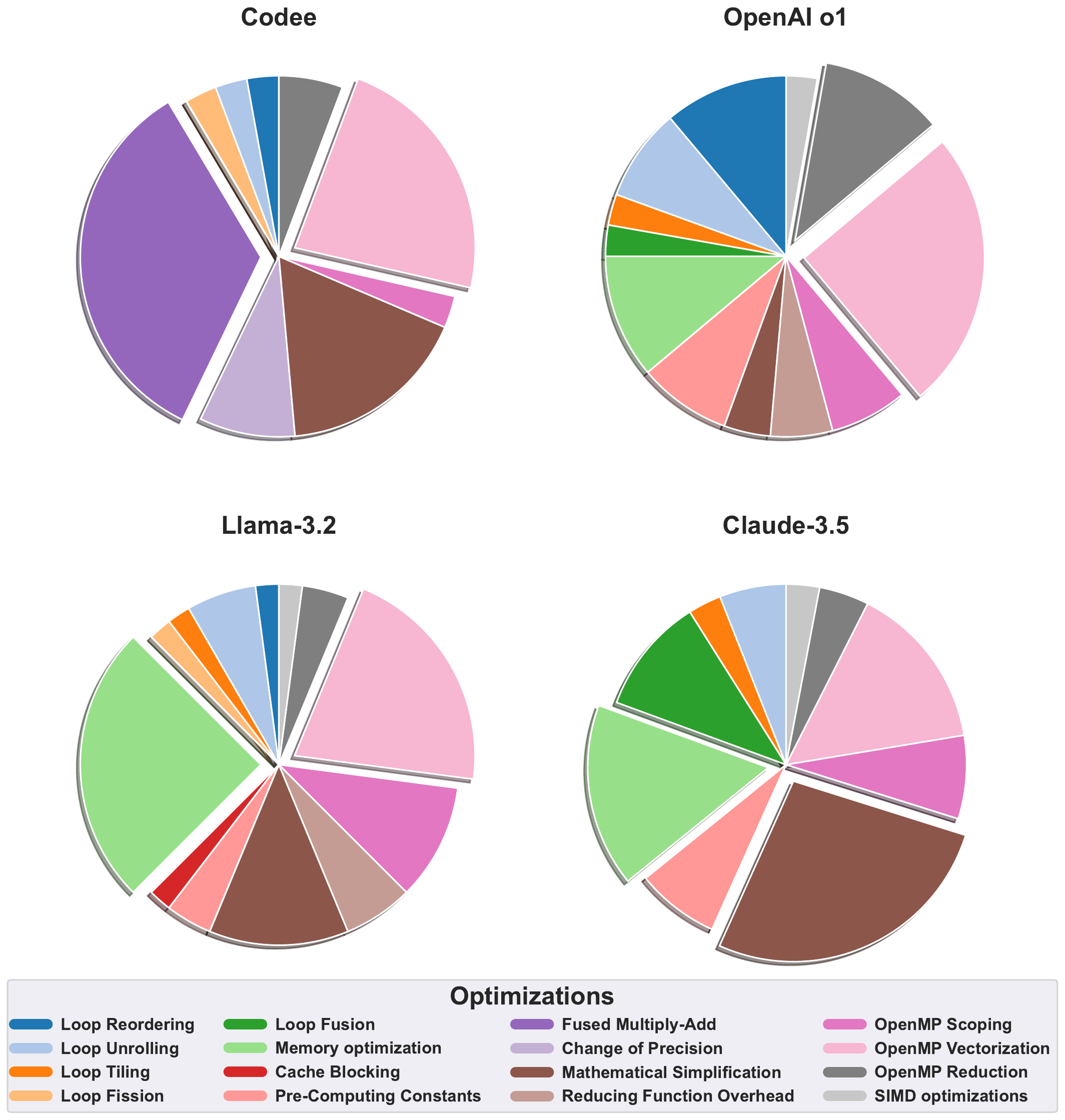}
  \caption{Distribution of optimizations applied by each tool}
\label{fig:EX6}
\vspace{-1.0em}
\end{figure}

%% file: src/CaseStudies.tex
\section{Case Studies}
\label{sec:case studies}
We describe the evaluation of level 3 benchmarks in Table~\ref{tab:dataset} in this section.
Experiments were performed on up to eight \textit{AMD EPYC 7543} processors.
We provide first an overview of an agent system we designed to conduct studies.
Subsequently, we discuss our findings on each benchmark.

\subsection{Performance Optimization Agent}

\input{figure/Agent_Diagram}

Developing and improving HPC applications is an intricate and iterative process that targets efficiency and scalability.
While traditional tools provide insights through runtime or static inspection, the interpretation and application of these insights rely heavily on human expertise.
This process can be time-intensive and error-prone.
Advancements in AI-driven code assistants \cite{rozière2024codellamaopenfoundation,guo2024deepseekcoderlargelanguagemodel,li2023starcodersourceyou} are bringing a paradigm shift in this domain.
On the other hand, while LLMs excel in automatically generating small and general-purpose code, they often fail to analyze large HPC codebases due to the lack of understanding of program structure, invocation context, and performance characteristics.

To address these challenges, we developed a performance optimization agent leveraging OpenAI o1 as the ``brain'' of the system, as illustrated in Figure~\ref{fig:Agent}.
This agent facilitates HPC code development by integrating profiling feedback into the optimization process.
It serves as a baseline for evaluating the effectiveness of LLMs in enhancing productivity in real HPC application development.
Similar to other agent systems, our performance optimization agent comprises core components: \textit{Memory}, \textit{Tools}, \textit{Planning}, and \textit{User Requests}. The agent's workflow begins with time-based profiling using HPCToolkit.
Profiling data is then converted into JSON structures---more suitable for LLM processing---using Hatchet~\cite{10.1145/3295500.3356219}.
In the profiling data, we include information on individual source lines, their invocation contexts, and associated metrics such as total CPU execution time and L1 data cache miss rate.
In addition to profiling data, user-provided inputs are incorporated to enrich the execution context.
These inputs include environment configurations (e.g., hardware and software resources), the number of threads, ranks, and iterations for the application's runtime.
With this comprehensive input, the LLM generates optimized code and recommends additional metrics for further performance analysis.
Following this, the identified hotspot function is replaced with its optimized version.
The agent then recompiles the application and measures performance improvements.
This optimization loop continues iteratively until a predefined threshold (e.g., three iterations) is reached or the LLM suggests no further optimizations.
Importantly, the metrics and program context obtained during each iteration are kept in the agent's memory to allow the LLM to reference prior iterations for more informed optimization suggestions.

\subsection{Case 1: NPB\_CG}

The \textit{Conjugate Gradient (CG)} benchmark, part of the \textit{NAS parallel Benchmarks (NPB)} suite, evaluates the performance of parallel supercomputers by simulating the solution of sparse linear systems using the conjugate gradient method.

The \textit{conj\_grad} function was identified as the hotspot using HPCToolkit.
Codee provided one L2 optimization suggestion by applying multithreading parallelism to the specific forall loop on the hotspot. 
The original code ran $25.00$ seconds, where the hotspot spent $22.00$ seconds ($88.1\%$).
The code optimized by Codee running with 16 threads took $4.58$ seconds ($5.46\times$).
To compare with, applying the performance optimization agent with multiple iterations using 16 threads yielded a speedup of up to $8.22\times$ with no discrepancies in output correctness. 

\paragraph{Version 1}

The agent optimized the code using \texttt{\#pragma omp parallel} to parallelize loops, \texttt{\#pragma unroll} to unroll loops, and \texttt{\#pragma omp simd} to enable vectorization. 
To be specific, in order to ensure the correctness for the CG method, it used  \texttt{\#pragma omp parallel for reduction(+: rho)} or \texttt{\#pragma omp parallel for reduction(+: sum)} for corresponding loops to avoid data races.
The optimized code's runtime decreased to $3.04$ seconds ($8.22\times$).


\paragraph{Version 2}

In this version, the agent provided two additional performance metrics: L1 cache misses and the number of floating-point instructions, hinting that effectively accessed data and utilized CPU vector units can lead to potential speedups.
The agent then applied compiler directives (e.g. \texttt{\_\_restrict}), OpenMP reductions, SIMD parallelism, and alignment hints to the specific loop.
Furthermore, it added OpenMP parallelization and prefetch instruction to the \texttt{conj\_grad} function, which was not parallelized in the previous version.
However, following the optimizations, the total runtime increased to $5.69$ seconds compared with the previous version.

\paragraph{Version 3}
Based on profiling results, L1 cache misses and floating-point instruction counts remained nearly unchanged, and the thread loads across the three indicators are highly imbalanced.
To this end, the agent applied advanced optimizations for \texttt{conj\_grad}, such as dynamic scheduling (\texttt{\#pragma omp parallel schedule}) for threads and aligning the addresses of arrays with \texttt{alignas(64)}.
The final runtime was decreased to $5.52$ seconds compared with version 2, but L1 cache misses remained the same as version 2.



\subsection{Case 2: XSBench}

\textit{XSBench} is a compact application that simulates a crucial computation kernel of the Monte Carlo neutron transport algorithm.
The hotspot is located in the \texttt{calculate\_macrox\_xs} function identified using HPCToolkit.
The original code was executed in 39.5 seconds, and the \texttt{calculate\_macrox\_xs} function took 28.3 seconds (71.5\%).
Codee didn't identify any optimization opportunities for this function.
 
Applying the performance optimization agent with three iterations yielded $1.33\times$ speedup.
Below we describe metrics profiled and optimizations applied by the agent.


\paragraph{Version 1}
In the first iteration, the agent applied the SIMD intrinsics from \texttt{<immintrin.h>} to optimize the hotspot function, suggesting it helped accelerate larger datasets.
Then, it employed compiler hints such as \texttt{\_\_attribute((hot,flatten,always\_inline))} to reduce the function invocation overhead because the hotspot function was called in the innermost level of a nested loop.
In addition, the agent also annotated \texttt{\_\_builtin\_expect} for certain conditions, aiming to improve branch prediction. 
Yet, the overall runtime slightly increased to 40.1 seconds.

\paragraph{Version 2}
By investigating L1 data cache load misses as a new profiling metric, the agent pointed out that memory access optimization could lead to potential speedups.
The agent reduced the L1 data cache load misses by adopting prefetch instructions (e.g. \texttt{\_\_bultin\_prefetch}) for frequently accessed pointers and introducing local variables for frequently accessed positions from \texttt{high} and \texttt{low} arrays.
Moreover, the agent duplicated the calculation of interpolation factors to reduce memory accesses. 
After these optimizations, the overall runtime decreased to $34.6$ seconds, and the hotspot runtime decreased to $20.2$ seconds. 


\paragraph{Version 3}
Lastly, the agent focused on leveraging SIMD intrinsics to further reduce L1 data cache load misses and improve performance, which is implemented by adopting \texttt{\_m256d} functions for vectorization.
It also unrolled the loops around the hotspot function.
Applying the above optimizations achieved the best result, with a runtime of 29.8 seconds.



\subsection{Case 3: LBM D2Q37}
\textit{LBM D2Q37} is a Computational Fluid Dynamics (CFD) code that simulates 2D fluids using the ``Lattice Boltzmann Method'' \textit{(LBM)} with 37 velocity components.
The computation is performed in double precision, featuring two key kernels: the memory-bound \texttt{propagate} kernel and the compute-bound \texttt{collide} kernel. 
We applied our performance agent to optimize LBM in single and multi-process environments using the ``Test'' dataset.

\subsubsection{Single Process}
The hotspot was identified in the \texttt{collide} kernel based on profiling data obtained using HPCToolkit.
Codee only suggested one optimization opportunity for the hotspot, indicating lower chances of achieving an increase in speedup.
It suggested applying optimizations like ``change of memory access pattern'' from column-major order to row-major order and applying either ``loop fission'', ``loop tiling'', or ``loop fusion'' to avoid non-consecutive/indirect array access. 
The original code was executed in 98.4 seconds, while the code with loop tiling ran slightly longer, taking 106 seconds, achieving a speedup < 1. 

Applying the performance optimization agent with multiple iterations also yielded no improvements in performance, with speedups \textasciitilde 1 and no discrepancies in output correctness.

\paragraph{Version 1}
The first set of optimizations applied by the agent are using OpenMP \texttt{\#pragma omp simd} directives to enable vectorization in the innermost loops over the hotspot.
These directives were added for loops that computed values like \texttt{rho}, \texttt{u}, \texttt{v}, \texttt{Hermite\_projections}, \texttt{Collisional\_updates}, and the final projections written to \texttt{nxt}.
The goal was to leverage data parallelism in these loops.
However, the total runtime only slightly decreased to 98.3 seconds.

\paragraph{Version 2}
The agent requested new profiling metrics to inspect L1 cache inefficiencies, suggesting memory access optimizations could yield potential speedups.
Based on the new metrics, the optimized code refined the placement of \texttt{\#pragma omp simd} to specific computational loops.
The runtime marginally increased to 98.6 seconds, with profiling metrics showing an increase in L1 data cache misses and stalled cycles, indicating the failure of the previous assumption that the placement of SIMD instructions causes inefficient memory access patterns.

\paragraph{Version 3}
Finally, to address the memory bottlenecks, more aggressive optimizations were applied by the agent, including compiler-specific directives like \texttt{\#pragma GCC ivdep} to ignore assumed dependencies and loop unrolling to improve throughput.
Additionally, \texttt{\_\_builtin\_prefetch} was used to prefetch data into cache, aiming to reduce L1 cache miss latency.
While these advanced optimizations were better aligned with profiling insights, they only achieved a runtime of 98.5 seconds, again yielding a speedup close to 1.


\subsubsection{Multiple Processes} The benchmark was profiled using eight MPI ranks.
Per rank performance metrics were obtained by the agent.
By observing that all processes were taking a long time in the same hotspot loop, the agent optimized the code by increasing parallelism by using four threads per rank, instead of two in the original code.
However, the total runtime rose from 9.7 seconds to 10.8 seconds.
The hotspot loop still consumed the majority of the total compute time, though it averaged at 85.7\% of total runtime instead of 91.7\%, indicating that some computational overhead outside of the hotspot had grown. 
The profiling data for each rank showed a spike in front-end stalls—often over 80\%—which implied the CPU pipeline was waiting for instruction fetch or decode.
This can occur when additional threads compete for the same instruction stream.
Another possible factor is the synchronization overhead and increased thread scheduling complexity.



\subsection{Case 4: Minisweep}
Minisweep is a radiation transport benchmark that reproduces the computational pattern of the sweep kernel of the Denovo Sn radiation transport code \cite{Evans01082010}. The sweep kernel, responsible for most of Denovo’s computational cost, models radiation transport for nuclear reactors.
We applied the performance agent to optimize Minisweep in a single-process environment using the ``Test'' dataset.

The hotspot was identified in the \texttt{sweeper\_kba\_c\_kernels.h} file based on profiling data obtained using HPCToolkit.
Codee failed to suggest meaningful optimizations for the hotspot itself.
Applying the performance optimization agent with multiple iterations also yielded no improvements, resulting in a similar runtime to the original code of 3.79 seconds.


\paragraph{Version 1}
\sloppy
The first set of optimizations applied
\texttt{\_\_attribute\_\_((always\_inline))} to enable full inlining of the \texttt{Sweeper\_sweep\_cell} function and inserted OpenMP \texttt{\#pragma omp simd} directives to enable vectorization in the innermost loops over the hotspot. Additionally, \texttt{\#pragma unroll} and \texttt{\#pragma ivdep} were used to improve throughput and allow for more aggressive compiler optimizations.
These changes aimed to leverage data parallelism. Despite this, the runtime only slightly decreased to 3.77 seconds.

\paragraph{Version 2}
In the second iteration, new profiling metrics including measurements of frontend and backend stalled cycles were applied.
The profiling data showed that the hotspot loop accounted for 55.5\% of the computation time, with 43.8\% stalled frontend cycles and 39.1\% stalled backend cycles. 
Based on the new metrics, the optimized code refined the placement of \texttt{\#pragma GCC ivdep} to ignore assumed dependencies, \texttt{\#pragma unroll} to improve throughput, and \texttt{\#pragma omp simd} to enhance vectorization.
These changes aimed to reduce frontend stalls by improving instruction flow and backend stalls by enhancing parallel execution efficiency.
However, the new version resulted in a slightly increased runtime of 3.79 seconds.

\paragraph{Version 3}
In the third iteration, the agent applied the same optimization as the previous version but refined the application of optimization directives.
The compiler-specific directives were inserted around the hotspot.
These changes led to a slightly increased runtime of 3.88 seconds.
Profiling metrics showed a significant decrease in backend stalls at 30.9\% indicating that while some improvements were made, the root causes of frontend instruction dependencies persisted.


%% file: figure/Agent_Diagram.tex
\begin{figure}[ht]
\centering
  \includegraphics[width=1\linewidth]{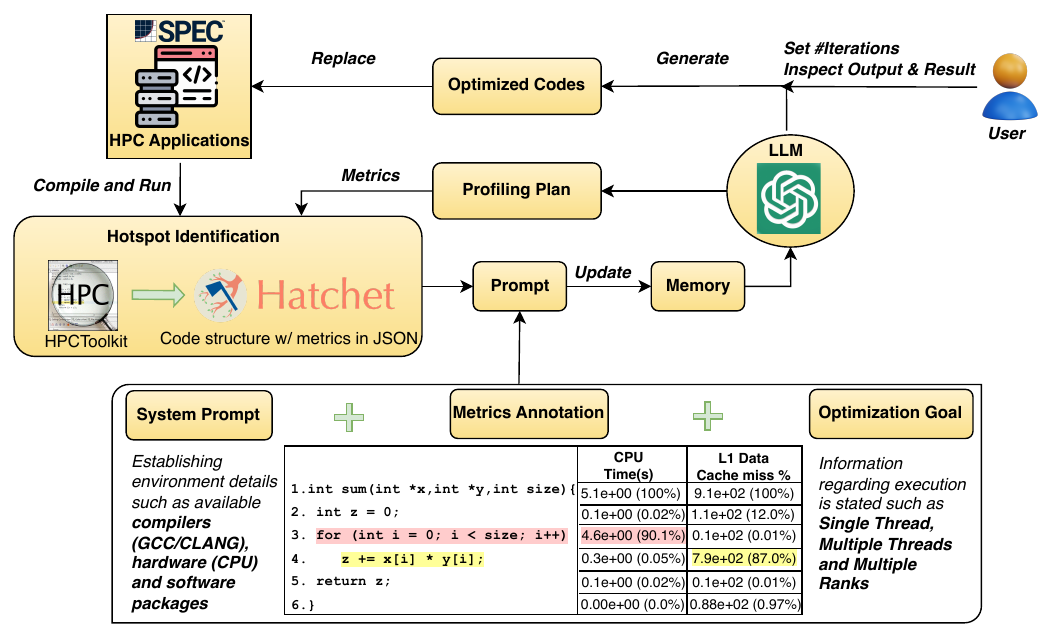}
  \caption{Overview of the Performance Optimization Agent}
\label{fig:Agent}
\vspace{-1.0em}
\end{figure}

%% file: src/Discussion.tex
\section{Discussion}
\label{sec:discussion}
\input{table/Star_rating}

We assess the overall performance of the tools employed in the studies by defining and evaluating five key metrics:

\begin{itemize}
    \item \textbf{Speedup:} Measures the average speed improvement achieved by the tools across various experiments. Higher rankings indicate greater speedups.
    \item \textbf{Correctness:} Evaluates the number of benchmarks for which the tools produced accurate results, with higher rankings reflecting greater reliability.
    \item \textbf{Commonsense:} Assesses the tools' effectiveness across multiple domains. Higher ranking indicates more domains a tool leads in terms of the average speedup.
    \item \textbf{Latency:} Measures the time required by each tool to apply optimizations. Higher rankings indicate a lower latency.
    \item \textbf{Applicability:} Examines the tools' ability to optimize larger benchmarks featured in the case studies.
\end{itemize}

Each tool is rated on a scale of 1-4 stars (4 being the best and 1 being the worst) for each of these metrics, as shown in Table~\ref{tab:tool_ratings}.
From our evaluation, Codee ranks as the best overall in optimizing HPC codes, with OpenAI o1 being a close second.
Claude-3.5 ranks third, while Llama-3.2 ranks last.
It is noteworthy that both Llama-3.2 and Claude-3.5 generate many incorrect codes across experiments; Claude-3.5 achieves a higher speedup in general.
Codee's strength lies in its capability to invoke reliable and accurate compiler-based analysis, exhibit low response latency, and avoid context length limitations.
Nonetheless, its primary limitation lies in the fact that the optimization patterns are predefined and lack the flexibility for customization according to user-specific code, and it is also incapable of offering algorithm-related enhancements.
Although LLMs generally fall short in rendering correct code, we observed that they have great potential in suggesting optimizations adaptive to user code.
Notably, OpenAI o1, as a reasoning model, performs quite well in optimizing code, especially in suggesting algorithm-related optimizations, yielding speedups even better than Codee. 
It is also worth noting that neither Codee nor LLMs have proven to be mature tools for automatically optimizing large-scale codebases.




%% file: table/Star_rating.tex
\begin{table}[h!]
\centering
\renewcommand{\arraystretch}{1.2} 
\setlength{\tabcolsep}{1.75pt} 
\small
\footnotesize
\begin{tabular}{|c|c|c|c|c|c|}
\hline
\textbf{Tool} & \textbf{Speedup} & \textbf{Correctness} & \textbf{Commonsense} & \textbf{Latency} & \textbf{Applicability} \\ \hline
Codee            & \threestars{}     & \fullstars{}        & \fullstars{}          & \fullstars{}     & \onestar{}          \\ \hline
OpenAI o1             & \fullstars{}      & \threestars{}         & \threestars{}        & \twostars{}      & \twostars{}        \\ \hline
Claude-3.5            & \twostars{}    & \twostars{}           & \threestars{}          & \threestars{}      & \textbf{NA}       \\ \hline
Llama-3.2             & \onestar{}       & \twostars{}          & \onestar{}         & \onestar{}    & \textbf{NA}          \\ \hline

\end{tabular}
\caption{Overall tool performance ranked by the total star count}
\label{tab:tool_ratings}
\end{table}

%% file: src/Conclusions.tex
\section{Conclusions} 
\label{sec:conclusions}

In this study, we designed and conducted a series of experiments using HPC-focused benchmarks to compare the impact of LLMs on performance optimization against state-of-the-art compiler-based tools.
Our evaluation reveals that state-of-the-art LLMs lag behind in several critical aspects, especially in terms of correctness.
However, we recognize the potential for integrating LLMs with traditional tools, as traditional tools typically offer only a subset of possible transformations.
Our performance optimization agent system also reveals a potential to integrate LLMs with profilers to circumvent its limitation on accessing the execution environment.
In the future, we plan to evaluate hardware performance not limited to CPUs.
Additionally, we aim to enhance our agent systems through the utilization of structured inputs, more concise prompts, and static code analysis.

